\documentclass[a4paper,10pt]{article}
\usepackage{authblk}
\usepackage[utf8]{inputenc}

\usepackage{url}
\usepackage{hyperref}
\usepackage{float}
\usepackage{amssymb}
\usepackage{amsmath}
\usepackage{array}
\usepackage{xcolor}
\usepackage{graphicx}
\usepackage{mathrsfs}

\DeclareMathOperator{\erf}{erf}
\DeclareMathOperator{\erfc}{erfc}
\DeclareMathOperator{\erfi}{erfi}

\usepackage{subfigure}
\graphicspath{{./images}}

\begin{document}

\title{\large\bf  Quantum Cosmology in Page-Wootters' Framework: Singularity Resolution and  Clock Coherence}
 
\author{\normalsize\bf Vishal\thanks{\tt vishal.phiitg@iitg.ac.in} \ and Malay K. Nandy\thanks{\tt mknandy@iitg.ac.in} \\ \normalsize\em Department of Physics, Indian Institute of Technology Guwahati \\ \normalsize\em Guwahati 781 039, India} 

\date{\normalsize\rm (9 October 2026)}

\maketitle  
\begin{abstract}
We investigate the fate of  big-bang singularity and coherence between distinct clock states in a quantum cosmological setting within the Page-Wootters framework. With a canonically quantized plane-symmetric Bianchi I universe, we  employ Dirac factorization of the Wheeler-DeWitt constraint and obtain a Schr\"odinger-like equation which is first-order in an internal clock variable. By constructing its solution as a Gaussian wave packet in the expanding branch, we  derive the corresponding conditional state and probability density of the universe. We find that the conditional probability density vanishes in the zero-volume limit, indicating  resolution of the big-bang singularity within this probabilistic framework. Another prespective of the study is provided by the construction of  density matrix from the conditional states associated with different clock readings and the remaining degree of freedom. By tracing out the remaining subsystem, we obtain the reduced density matrix and analyze the resulting coherence as a function of  separation between different clock states. These results provide a relational characterization of singularity resolution and demonstrate  how the  spectral properties of the quantum state govern the  temporal coherence between different clock states in the Page-Wootters description.\\

\end{abstract} 

\tableofcontents

\section{Introduction}
\label{introduction}
The fundamentally different roles played by time in general relativity and quantum mechanics give rise to a central conceptual difficulty in the attempt to reconcile the two theories within a quantum theory of gravity. In general relativity, time is not an external parameter but is intertwined with the dynamical geometry of spacetime~\cite{landau1971classical}, whereas in  quantum mechanics it enters as an external parameter with respect to which the quantum state evolves~\cite{dirac1981principles}. This distinction becomes particularly pronounced in the canonical approach to quantum gravity~\cite{WheelerQG,dewitt1967quantum}. Upon quantization of the Hamiltonian constraint, one obtains the Wheeler-DeWitt equation, which contains no explicit external time parameter and takes the form of a constraint on the physical state of the universe. Consequently, the fundamental quantum description is formally timeless, raising the question of how dynamical evolution and temporal ordering can emerge from a theory in which time does not appear explicitly. This conceptual difficulty give rise to the well-known ``problem of time" in quantum gravity~\cite{unruh1989time,kuchavr2011time,isham1993canonical}.

Since the Wheeler-DeWitt equation does not single out a preferred notion of time, one must determine which degree of freedom, if any, can serve as a temporal reference. This leads to related difficulties of defining physical observables and specifying the appropriate Hilbert space of physical states~\cite{feinberg1988}. A central question is therefore when and how a notion of time should be introduced: it may be selected at the classical level and retained during quantization, identified through a suitable choice of variables within the quantum theory, or regarded as an emergent concept arising from correlations within the quantum state itself. The choice of time is also closely connected with the probabilistic interpretation of the theory, since the usual formulation of quantum probabilities is based on a well-defined evolution parameter and a corresponding conserved inner product~\cite{anderson2012problem}.

The problem of time becomes particularly significant in quantum cosmology, where the universe itself is treated as a closed quantum system and no external temporal reference is available~\cite{hartle2014spacetime}. In the absence of an external or preferred notion of time, the description of quantum dynamics in the early universe becomes conceptually nontrivial, since the evolution of the quantum state must be formulated entirely in terms of degrees of freedom internal to the universe. This issue is especially relevant to the study of the classical big bang singularity~\cite{kiefer_zeh1995,Ashtekar_2011}, where the spacetime description of general relativity breaks down and a quantum description of the gravitational degrees of freedom is expected to become essential.

Singularity resolution has been extensively investigated within quantum cosmology, including in the Wheeler--DeWitt framework, where different choices of boundary conditions and minisuperspace models have been explored in connection with the fate of the big bang singularity~\cite{misner1969quantum,hartle1983wave}. Subsequent works  suggested  that quantum effects can significantly modify the classical cosmological dynamics, giving rise to phenomena such as quantum bounces and oscillatory behavior~\cite{bojowald2001,ashtekar2006,bouhmadi2014resolution,kiefer2019singularity}.

While these works among others highlight  that a quantum description of the early universe must address  the fate of the big bang singularity, it is also essential to address the  problem of absence of an external time parameter inherent in the quantum cosmological  formulation.  In particular, it becomes important to answer how physical probabilities and dynamical evolution can be defined so as to  assess the big bang singularity in the quantum  universe.

These considerations suggest that, in quantum cosmology, time should not be regarded as an external parameter imposed independently of the quantum system, but should instead be identified from the intrinsic dynamical degrees of freedom and their correlations within the system~\cite{rovelli1991time,Altaie2022pwh}. A particularly relevant perspective is the relational approach~\cite{rovelliQM,hohn2021trinity}, in which physical evolution is described through correlations between observables, is worth mentioning. 
In such an approach,  one degree of freedom is used as an internal reference, while the behavior of the remaining degrees of freedom is described relative to it.

Related ideas arise in other timeless formulations, such as the consistent-histories approach~\cite{griffiths1984consistent,gell1996quantum}, where physical predictions are formulated in terms of sequences of alternatives and their associated probabilities rather than states evolving with respect to an external time parameter. These approaches demonstrate that dynamical and probabilistic descriptions can be constructed intrinsically from the quantum system through correlations and histories.

A prominent framework for describing relational dynamics in a closed quantum system is the Page-Wootters (PW) formalism~\cite{page1983evolution}. In this approach, the global state of the closed system is stationary and satisfies the Hamiltonian constraint, consistent with the absence of an external time parameter. Temporal evolution is instead recovered relationally by decomposing the total system into a clock subsystem and a subsystem consisting of the remaining degrees of freedom.
The conditional probability of the remaining subsystem, to be found in a particular state corresponding to a clock reading, is focused upon.  In this way, the apparent evolution of the system is embedded in the quantum correlations between the clock and the remaining degrees of freedom, rather than in an evolution with respect to an external time parameter~\cite{Gambini_2004,gambini2009conditional,Giovannetti2015,mv2017}.

Beyond the conditional probability itself, however, the relational description contains information about the quantum correlation between different clock readings that is not captured by a single conditional state~\cite{Moreva2017zhw}. In particular, different clock readings correspond to different conditional states of the remaining subsystem, and it is natural to ask how strongly these relational descriptions remain quantum mechanically coherent as the clock separation increases.

This question can be investigated through the reduced density matrix of the clock subsystem, obtained by tracing over the remaining degrees of freedom. Its non-diagonal elements in the clock basis quantify the coherence between different clock readings and are directly related to the overlap of the corresponding conditional states.

The question of quantum coherence is closely connected with the broader study of decoherence in quantum cosmology. Decoherence refers to the suppression of interference between different components of a quantum state as a result of interaction with degrees of freedom that are not included in the effective description. It has played an important role in attempts to understand how classical description of the universe can emerge from an underlying quantum state~\cite{padmanabhanDecoherence1989}. In quantum cosmology, gravitational perturbations, matter degrees of freedom, or other modes of the universe can provide an effective environment for a selected set of variables, leading to a reduced description in which interference between sufficiently distinct alternatives is suppressed~\cite{halliwellDecoherence1989, LaFlamme1990kd, barvinskyDecoherence1999}. Such analyses attempt to provide a connection between the quantum cosmological states and the classical configuration of the universe.

In this work, we investigate the big bang singularity within the Page-Wootters framework and examine the quantum coherence between distinct clock states. We consider the canonically quantizatized  plane-symmetric Bianchi I universe. Employing Dirac decomposition to the Wheeler-DeWitt constraint, we obtain a Schrodinger-like equation which is first-order in an internal clock variable. We construct its solution as  a wave packet in the momentum space. 
This construction facilitates in obtaining the corresponding conditional states and probabilities for describing the cosmological dynamics relationally.

We find that the conditional probability density vanishes in the zero-volume limit, indicating resolution of the big bang singularity within this probabilistic framework. Another prespective of the study is provided by the construction of  density matrix from the conditional states associated with different clock readings and the remaining degree of freedom. By tracing out the remaining subsystem, we obtain the reduced density matrix and analyze the resulting coherence as a function of  separation between different clock states. This analysis provides a quantitative characterization of the dependence of  temporal coherence on the spectral properties of the quantum state.

The remainder of the paper is organized as follows. Section~\ref{pw-framework} introduces the Page-Wootters formalism. Section~\ref{Bianchi_universe} presents the Wheeler-DeWitt equation for the plane-symmetric Bianchi I universe and introduces the clock variable within  Page-Wootters framework. In Section~\ref{singularity}, we construct conditional states and probabilities and examine the behavior of the quantum state in the vicinity of the big bang singularity. Section~\ref{density_matrix_coherence} is devoted to the construction of reduced density matrix and the analysis of coherence between different clock states. Finally, Section~\ref{discussion} concludes the paper with a discussion.

\section{The Page-Wootters Formalism}
\label{pw-framework}
The Page-Wootters (PW) formalism~\cite{page1983evolution}  offers  a suitable  approach to the ``problem of time" in quantum cosmology by treating time as an internal quantum degree of freedom, represented by a clock subsystem
within an overall timeless state of the universe~\cite{emergence2023,rijavec2025}.
The physical state of a quantum gravitational system satisfies the Wheeler-DeWitt constraint,
\begin{equation}
\hat{H}|\Psi\rangle = 0.
\label{wd}
\end{equation}

This equation expresses the global stationarity of the universe and signifies the timeless nature of the total quantum state. Relational dynamics is obtained by choosing an internal degree of freedom to serve as a clock. The total Hilbert space of the universe $\mathcal{H} $ can be factorized into a clock and the remaining subsystem,
\begin{equation}
\mathcal{H}=\mathcal{H}_C\otimes\mathcal{H}_R,
\end{equation}
where $\mathcal{H}_C$ denotes the Hilbert space associated with the clock  and $\mathcal{H}_R$ represents the remaining part of the universe.
The condition for a feasible clock is that it should not interact with the rest of the universe so that the Hamiltonian can be written as,
\begin{equation}
\hat{H} = \hat{H}_C \otimes \mathbb{I}_R + \mathbb{I}_C \otimes \hat{H}_R,
\label{hamiltonian}
\end{equation}
where $\hat{H}_C$ and $\hat{H}_R$ act only on the individual Hilbert spaces of clock and remaining subsystem, respectively.

The clock is characterized by an observable $T_C$ conjugate to $H_C$, such that $[T_C, H_C] = i$, with eigenstates $| t \rangle_C$ satisfying $T_C | t \rangle_C = t | t \rangle_C$, where $t \in \mathbb{R}$ labels clock readings. The entanglement between the clock and the remaining subsystem establishes correlations between clock readings and the states of the subsystem, thereby providing a basis for the emergence of
relational dynamics~\cite{mv2017,rijavec2023}. The state of the universe can be expanded as
\begin{equation}
| \Psi \rangle = \int dt \, c(t) \, | t \rangle_C \, |\psi_t \rangle_R,
\label{eq:entangled}
\end{equation}
providing the dynamical evolution of the universe, with $|\psi_t \rangle_R$ the state of the remaining subsystem. Henceforth, we shall denote $|t \rangle_C$ by  
$|t \rangle$ for brevity.

The total density operator is expressed as $\rho = |\Psi\rangle \langle \Psi|$. For a given clock reading $t$, the density operator of the remaining subsystem $R$ can then be obtained through Everett's relative-state construction~\cite{evrett1957},
\begin{equation}
\rho_R (t) = \frac{ \mathrm{Tr}_C \left[ (|t\rangle\langle t| \otimes \mathbb{I}_R) \rho \right]}{\mathrm{Tr} \left[ (|t\rangle\langle t| \otimes \mathbb{I}_R) \rho\right]} = \  | \psi_t \rangle_R \, _R\langle \psi_t|.
\label{eq:relative}
\end{equation}

The conditioning on the clock reading $t$ does not require a physical measurement of the clock; rather, it specifies the state of the subsystem $R$ relative to the clock degree of freedom. The conditional state $\rho_R(t)$ therefore provides the basis for making physical predictions about observables of subsystem $R$ at a given clock reading $t$. Let $|\phi\rangle$ denote an eigenstate of an observable $\hat{O}$ acting on the subsystem $R$. The conditional probability of obtaining the value $\phi $ when the clock reads $t$ is given by
\begin{equation}
P(\langle \phi|t \rangle) = \frac{\mathrm{Tr} \left[ (|t\rangle\langle t| \otimes |\phi \rangle\langle \phi|) \rho \right]}{\mathrm{Tr} \left[ (|t\rangle\langle t| \otimes \mathbb{I}_R) \rho \right]}.
\label{conditional-probability}
\end{equation}

This formulation shows that physical predictions in the Page-Wootters framework are formulated relative to the clock reading, with the clock  providing a reference for the remaining subsystem. Even though the total state $|\Psi\rangle$ is
stationary, the conditional state $|\psi_t\rangle_R$  exhibits
non-trivial dynamics. In this way, the usual notion of quantum evolution emerges relationally from the correlations inherent in the global timeless state.

\section{Bianchi I Universe and Clock Variable}
\label{Bianchi_universe}

\subsection{Classical Framework}
Bianchi universes \cite{landau1971classical,ellis1999cosmological,misner1972magic} offer tractable anisotropic homogeneous models of the classical universe. In particular,  the  Bianchi I framework provides the simplest anisotropic metric, 
\begin{equation}
ds^2 = - N^2 dt^2  +a_1^2(t) (dx^2 +  dy^2) + a_2^2(t) dz^2,
\label{metric}
\end{equation}
with two spatial directions expanding or contracting identically whereas the third evolving independently with scale factors $a_1(t)$ and $a_2(t)$, respectively.

For the spatial part, the determinant  of the spatial metric is $\sqrt{h} = a_1^2 a_2$,  and Ricci scalar ${}^3\!R = 0$. Consequently, the  extrinsic curvature and its determinant are $K_{ij}= -\frac{\dot{h}_{ij}}{2N} =\frac{1}{N} \text{diag}(-a_1 \dot{a_1}, -a_1 \dot{a_1}, -a_2 \dot{a_2})$   and  $K   = \frac{-1}{N} (\frac{2 \dot{a_1}}{a_1} +\frac{\dot{a_2}}{a_2} )$, so that the ADM action is given by
\begin{equation}
S = \frac{1}{16 \pi G} \int dt \, d^3 x \; N \sqrt{h} \;({}^3\!R + K^{ij} K_{ij} - K^2)=  \frac{3}{4 \pi G} \int dt \, d^3x  \frac{e^{3 \alpha}}{N}  \left( \dot{\beta}^2-\dot{\alpha}^2 \right),
\label{action}
\end{equation}
upon expressing it in terms of Misner variables, $\alpha=\frac13\ln(a_1^2 a_2)$ and $\beta=\frac13\ln\!\left(\frac{a_2}{a_1}\right)$.

Upon setting $\frac{3}{4 \pi G}= 1$ and choosing the gauge $N= e^{3 \alpha}$, the Lagrangian of the system is $L =  \dot{\beta}^2-\dot{\alpha}^2$, so that the  Hamiltoniam constraint turns out to be
\begin{equation}
H = p^{2}_{\beta} - p^{2}_{\alpha}=0,
\label{Hamiltonian}
\end{equation}
where $p_\alpha$ and $p_\beta$ are the canonical momenta conjugate to the variables $\alpha$ and $\beta$, respectively.

\subsection{Wheeler-DeWitt Equation and Dirac Decomposition}
We consider this anisotropic model~\eqref{metric} in the minisuperspace representation of the Wheeler-DeWitt framework~\cite{WheelerQG,dewitt1967quantum} to study the behavior of the universe in the early quantal stage of its evolution.

Consequently, we quantize the system  canonically by promoting the classical momenta in the Hamiltonian constraint~\eqref{Hamiltonian} to quantum operators, $p_{\beta} \rightarrow -i\hbar\frac{\partial}{\partial \beta}$ and $p_{\alpha} \rightarrow -i\hbar\frac{\partial}{\partial \alpha}$, leading to the Wheeler-DeWitt equation,
\begin{equation}
\left[
\frac{\partial^2}{\partial \alpha^2}
-\frac{\partial^2}{\partial \beta^2}
\right] \Psi(\alpha, \beta) = 0.
\label{wdeq}
\end{equation}

Since this equation has a Klein-Gordon-type structure, the inner product is not positive definite in general on the full solution space. Consequently, this structure presents a difficulty for the probabilistic interpretation of the quantum state as the conserved current associated with Klein-Gordon equation  can become negative. It is therefore useful to recast the second-order Wheeler-DeWitt equation~\eqref{wdeq} into a first-order form, in which the positive  and negative frequency sectors can be treated separately. We achieve this through Dirac decomposition of the Wheeler-DeWitt constraint~\cite{craigsingh2010}.

Rewriting equation~\eqref{wdeq} by isolating the $\beta$-derivative, we obtain
\begin{equation}
-\frac{\partial^2 \Psi(\alpha,\beta)}{\partial \beta^2}
= \hat{\Omega} \, \Psi(\alpha,\beta),
\label{wd_omega}
\end{equation}
where  $\hat{\Omega} = \hat{p}_{\alpha}^{\,2}
= -\frac{\partial^2}{\partial \alpha^2}$. Dirac decomposition of equation~\eqref{wd_omega} consists of writing it in the first order form,
\begin{equation}
-i \frac{\partial \Psi(\alpha,\beta)}{\partial \beta} = \sqrt{\hat{\Omega}} \, \Psi(\alpha,\beta).
\label{WD_Dirac}
\end{equation}

\subsection{Entangled States and Clock Variable}
To apply the Page-Wootters relational formalism, we choose the
anisotropy variable $\beta$ as the clock degree of freedom and the volume variable $\alpha$ as the remaining subsystem. Consequently, the total Hamiltonian of the universe  splits into the clock and  rest of the universe, $\hat{H}_C = \hat{p}_\beta $ and $\hat{H}_R = \sqrt{\hat{\Omega}}$, respectively. The clock variable is canonically conjugate to the clock Hamiltonian $\hat{H}_C$,  
satisfying the commutation relation, $[\hat{\beta}, \hat{H}_C] = i$.

The eigenfunctions of $\hat{\Omega}$ are plane waves, $a_k(\alpha) = \frac{1}{\sqrt{2\pi}} e^{i k \alpha}$,  which satisfy $ \hat{\Omega} \, a_k(\alpha) = k^2 \, a_k(\alpha)$ as well as  $\sqrt{\hat{\Omega}} \, a_k(\alpha) = k \, a_k(\alpha)$. Restricting to positive frequencies $\omega = |k|$ driving the time evolution $e^{i |k| \beta}$, a general solution of equation~\eqref{WD_Dirac} can be written as
\begin{equation}
\Psi(\alpha,\beta) = \frac{1}{\sqrt{2\pi}} \int_{-\infty}^{\infty} dk \; F(k)\,
e^{i k \alpha} e^{i |k| \beta},
\label{psi_entangle}
\end{equation}
where $F(k)$ is the spectral decomposition of the wavefunction in the momentum space $k$. This superposed state represents an entanglement between the clock  and the remaining subsystem.

We write the total wavefunction as a combination of the left-moving (contracting) and right-moving (expanding) branches,
\begin{equation}
\Psi(\alpha,\beta) = \frac{1}{\sqrt{2\pi}} \int_{-\infty}^{0} dk \; F_-(k)\, e^{i k (\alpha - \beta)} + \frac{1}{\sqrt{2\pi}} \int_{0}^{\infty} dk \; F_+(k)\, e^{i k (\alpha + \beta)}.
\label{exp_contrac}
\end{equation} 
The functions $F_+(k)$ and $F_-(k)$ represent the amplitudes of the positive ($k>0$) and negative ($k<0$) momentum sectors, respectively. In general, these two functions are independent and characterize the relative contribution of the two branches of the wavefunction. 

In this work, we focus on the expanding branch $(k<0)$. The momentum eigenstates in the clock sector is defined through
\begin{equation}
\langle \beta | k \rangle_C = \frac{1}{\sqrt{2\pi}} e^{-i k \beta},
\end{equation}
and satisfy the orthonormality condition
\begin{equation}
\langle \beta | \beta' \rangle =\int dk \, \langle \beta | k \rangle_C \ {}_C\langle k | \beta' \rangle =\frac{1}{2\pi} \int dk \, e^{-ik(\beta-\beta')} = \delta(\beta' - \beta).
\end{equation}
Thus, the clock states corresponding to distinct values of $\beta$ are orthogonal and hence represent distinguishable clock readings. In addition, the vanishing commutator 
$[\hat{H}_C , \hat{H}_R ]=0$ ensures that the clock and the remaining subsystem do not interact dynamically. Furthermore, the total global state $|\Psi \rangle$ represents an etangled structure between the clock and the remaining susbsytem.

The above properties establish $\beta$ as a suitable clock variable within the Page-Wootters construction. The orthogonality of the clock states guarantees distinguishable clock readings, while the absence of direct interaction with the remaining subsystem preserves its role as a reference degree of freedom. In addition, the entangled structure of the global state establishes correlations between the clock and the remaining subsystem, which is essential for the construction of conditional states and the emergence of relational dynamics~\cite{Giovannetti2015,mv2017,moreva2014}.

\section{Relational Dynamics and Singularity Resolution}
\label{singularity}
In this section, we investigate the fate of big bang singularity in a relational setting using  Page-Wootters formalism.  We construct the conditional state of the remaining subsystem $\alpha$ for a given clock reading $\beta_0$ and subsequently derive the corresponding conditional probability density. This allows us to examine how the probability of different $\alpha$ configurations  depends on the clock reading $\beta_0$ and, in particular, to investigate the behavior of the conditional probability in the classical singular limit
$\alpha\rightarrow-\infty$.

\subsection{Construction of Conditional State}
Focusing exclusively on the expanding branch of the solution~\eqref{exp_contrac}, with $F_+(k) = 0$, the wavefunction  reduces to
\begin{equation}
\Psi(\alpha,\beta)
=
\frac{1}{\sqrt{2\pi}}
\int_{-\infty}^{0} dk \; F_-(k)\, e^{i k (\alpha - \beta)}.
\label{psi_exp}
\end{equation}

We choose a Gaussian profile for the momentum-space amplitude, $F_-(k) =  e^{-\frac{(k - k_0)^2}{2\sigma^2} }$, equation~\eqref{psi_exp} leads to the  wave packet
\begin{equation}
\Psi(\alpha,\beta) = \frac{1}{\sqrt{2\pi}} \int_{-\infty}^{0} dk\, \exp \left[
-\frac{(k-k_{0})^{2}}{2\sigma^{2}} + ik(\alpha-\beta) \right].
\label{wavepacket}
\end{equation}
Carrying out  the integration, we finally obtain (see Appendix~\ref{wavefunction-integral}) 
\begin{equation}
\Psi(\alpha,\beta) = \frac{ \sigma}{2} e^{i k_0 (\alpha-\beta)
- \frac{\sigma^2}{2}(\alpha-\beta)^2} \left[ 1 + \erf \left( \frac{-k_0 - i\sigma^2(\alpha-\beta)}{\sqrt{2}\sigma} \right) \right].
\label{wavefunctionfinal}
\end{equation}
The error function  arises due to the restriction of the momentum integration to the negative half-line. In the limit $-k_0 \gg \sigma$, the real part of the argument of the error function becomes large and positive, implying $\erf(z) \to 1$, and the wavefunction reduces to a standard Gaussian wavepacket. For finite values of $k_0$, however, the error function modifies the behavior  in the tail and leads to deviation from a purely Gaussian profile.

Since $\beta$ is choosen as the time variable, the normalized wavefunction  is defined as
\begin{equation}
\langle \Psi_N | \Psi_N \rangle = \int_{-\infty}^{\infty} \left|\Psi_N(\alpha, \beta)\right|^2\,d \alpha =1.
\label{normalization}
\end{equation}

We rewrite  equation~~\eqref{wavefunctionfinal} with the variables $x=\alpha-\beta, a=-\frac{k_0} {\sigma\sqrt{2}}$ and  $b=\frac{\sigma}{\sqrt{2}}$ so that the normalized wavefunction takes the compact form
\begin{equation}
\Psi_N(x) =  \frac{N \sigma}{2} e^{ik_0x} e^{-\frac{\sigma^2x^2}{2}}
\left[1 + \erf(a-ibx) \right]
\label{compactpsi}
\end{equation}
where $N$ is the normalization constant. Consequently, the normalization condition~\eqref{normalization} becomes
\begin{equation}
\frac{N^2 \sigma^2}{4} \int_{-\infty}^{\infty} d\, x \, e^{-\sigma^2 x^2} \left[1+\erf(a-ibx)\right] \left[1+\erf(a+ibx)\right] =1. 
\label{normintegral}
\end{equation}
Using the solution of this integral (see appendix~\ref{norm_calc}), we obtain  the normalization constant 
\begin{equation}
N = \left[\frac{2}{\sigma \sqrt{\pi} (1+\erf(-\frac{k_0}{\sigma}))}\right]^{\frac{1}{2}}.
\label{norm_constant}
\end{equation}

Having identified \(\beta\) as the clock variable, we now proceed to construct the conditional state corresponding to a definite clock reading \(\beta = \beta_0\). This is achieved by projecting the total wavefunction onto the clock eigenstate \(|\beta_0\rangle\), following the Page-Wootters prescription.
The global state within the present model can be expressed as
\begin{equation}
|\Psi_N \rangle = \int d\alpha \, d\beta \ \Psi_N(\alpha,\beta)\, |\beta\rangle \otimes |\alpha\rangle_R .
\label{global_state}
\end{equation}

To extract the relational dynamics, we project the total state onto a definite clock state $|\beta_0\rangle$. This is implemented by applying the projection operator $|\beta_0\rangle \langle \beta_0| \otimes \mathbb{I}_R$, yielding
\begin{equation}
|\beta_0\rangle \langle \beta_0| \otimes \mathbb{I}_R |\Psi_N \rangle 
= \int d\alpha \, d\beta \ \Psi_N (\alpha,\beta)\, |\beta_0\rangle \langle \beta_0 | \beta \rangle \otimes |\alpha\rangle_R.
\end{equation}

Using the orthonormality condition $\langle \beta_0 | \beta \rangle = \delta(\beta - \beta_0)$, the integration over $\beta$ can be performed immediately, leading to
\begin{equation}
|\beta_0\rangle \langle \beta_0| \otimes \mathbb{I}_R |\Psi_N\rangle 
= |\beta_0\rangle \otimes \int d\alpha \ \Psi_N(\alpha,\beta_0)\, |\alpha\rangle_R.
\label{projection}
\end{equation}
We thus identify the conditional state of the remaining subsystem at clock time $\beta_0$ as
\begin{equation}
|\psi_{\beta_0}\rangle_R = \langle \beta_0| \otimes \mathbb{I}_R |\Psi_N\rangle  = \int d\alpha \ \Psi_N(\alpha,\beta_0)\, |\alpha\rangle_R .
\label{condtn_state}
\end{equation}

Corresponding to~\eqref{condtn_state}, and using Eq.~\eqref{compactpsi} and~\eqref{norm_constant}, the conditional wavefunction is  given by 
\begin{equation}
\Psi_N(\alpha,\beta_0) = \frac{N \sigma}{2} e^{i k_0 (\alpha-\beta_0)
- \frac{\sigma^2}{2}(\alpha-\beta_0)^2} \left[ 1 + \erf \left( \frac{-k_0 - i\sigma^2(\alpha-\beta_0)}{\sqrt{2}\sigma} \right) \right].
\label{Norm_wavefunction}
\end{equation}

The nontrivial content of this construction is enrooted in the entanglement between the clock and the remaining degree of freedom. This quantum correlation associates different states of the remaining subsystem  conditioned on a clock reading, thereby giving rise to an effective notion of relational dynamics.

In this sense, the parameter $\beta_0$ does not represent an external time variable, but rather labels different conditional states of the remaining subsystem relative to the clock. Having obtained the conditional state, we can now define the corresponding conditional probability density for the variable $\alpha$, which allows us to extract physically meaningful predictions within this relational framework.

\subsection{Conditional Probability of the Big Bang Singularity}
The conditional probability density $P(\alpha|\beta_0)$ with respect to the volume parameter $\alpha$ is defined as
\begin{equation}
P(\alpha | \beta_0) = |\Psi_N(\alpha, \beta_0)|^2.
\end{equation}
Using the conditional wavefunction from Eq.~\eqref{Norm_wavefunction}, we obtain the conditional probability density for the variable $\alpha$ at a fixed clock reading $\beta_0$ as
\begin{equation}
P(\alpha|\beta_0) = \frac{N^2 \sigma^2}{4} e^{-\sigma^2(\alpha-\beta_0)^2} \left[ 1+ \erf \left( \frac{-k_0-i\sigma^2(\alpha-\beta_0)}
{\sigma\sqrt{2}} \right) \right] \left[ 1+ \erf \left( \frac{-k_0+i\sigma^2(\alpha-\beta_0)} {\sigma\sqrt{2}} \right) \right].
\label{conditional_probability_1}
\end{equation}

To examine the behaviour of the conditional probability density~\eqref{conditional_probability_1} in the vanishing volume limit $\alpha\rightarrow-\infty$ for a finite clock reading $\beta_0$, we introduce the variables, $a=-\frac{k_0}{\sigma\sqrt{2}}>0$, and $Y=\frac{\sigma(\alpha-\beta_0)}{\sqrt{2}},$
so that the arguments of the error function can be written as $z_{\pm}=a\pm iY.$ In terms of these variables, equation~\eqref{conditional_probability_1} can be written as 
\begin{equation}
P(\alpha|\beta_0) = \frac{N^2 \sigma^2}{4} e^{-2Y^2} \left[1+\erf(z_-)\right] \left[1+\erf(z_+)\right].
\label{P_Y_form}
\end{equation}
Since the constant $a$ is finite, in the limit $\alpha\rightarrow-\infty$ with a finite clock reading $\beta_0$, we have $|Y| \to \infty$, giving $|z_{\pm}| \to \infty$ and  $|\arg z_{\pm}|<\frac{3\pi}{4}$. Thus, the standard large-$|z|$ asymptotic expansion of the complementary error function can be employed~\cite{abramowitz1964handbook}:
\begin{equation}
\erfc(z) = \frac{e^{-z^2}}{\sqrt{\pi}\,z} \left[ 1-\frac{1}{2z^2} +\frac{3}{4z^4}
-\frac{15}{8z^6} +\cdots \right], \qquad |z|\rightarrow\infty.
\label{erfc_asymptotic}
\end{equation}

Employing this expansion to the  conditional probability density~\eqref{P_Y_form} (see Appendix~\ref{probability_analysis}), we obtain  
\begin{align}
P(\alpha|\beta_0) = \;& N^2 \sigma^2 e^{-2Y^2} -\frac{N^2 \sigma^2}{2 \sqrt{\pi}}
\frac{e^{-a^2-Y^2}e^{2iaY}} {a-iY} \left( 1-\frac{1}{2(a-iY)^2} +\cdots \right)
\nonumber\\
&-\frac{N^2 \sigma^2}{2 \sqrt{\pi}} \frac{e^{-a^2-Y^2}e^{-2iaY}} {a+iY} \left(
1-\frac{1}{2(a+iY)^2} +\cdots \right) \nonumber\\
&+ \frac{N^2 \sigma^2 e^{-2a^2}} {4 \pi(a^2+Y^2)} \Bigg[ 1 -\frac{a^2-Y^2}{(a^2+Y^2)^2}
+\frac{1}{4(a^2+Y^2)^2} +\cdots \Bigg].
\label{P_complete_asymptotic_Y}
\end{align}

For $|Y|\to \infty$,  first term is exponentially suppressed, $ e^{-2Y^2} \to 0$, whereas the second and third terms behave as $\frac{e^{-Y^2}}{|Y|} \to 0$. The fourth term gives the  contribution, $
\frac{1} {Y^2} \to 0.$ Since $Y \to \infty$ and $\alpha \to -\infty$ are equivalent limits, the  conditional probability density  vanishes in this limit, 
\begin{equation}
\lim_{\alpha\rightarrow-\infty}
P(\alpha|\beta_0)=0, 
\label{singularity_avoidance_conditional}
\end{equation}
for every finite clock reading $\beta_0$.

This result implies that configurations corresponding to vanishing volume, $\alpha =\frac13\ln v \to -\infty$, that is $v= a_1^2 a_2 \to 0,$ are suppressed at the level of conditional probabilities for all clock values.  Since this limit corresponds to the classical big bang singularity at $v=0$, the vanishing of the conditional probability at $\alpha\to-\infty$ provides evidence for singularity avoidance within the Page-Wootters framework. Thus, the emergence of relational dynamics, driven by the correlation between the clock and the remaining subsystem, leads to a nonsingular description of the quantum universe.

\section{Density Matrix and Clock Coherence}
\label{density_matrix_coherence}
The conditional probability~\eqref{conditional_probability_1} describes the state of the system relative to a \emph{single} clock reading. It is therefore natural to ask a further question: how different are the quantum states of the system associated with two different readings of the clock, say $\beta_0$ and $\beta_0'$? This question concerns the correlation between the clock and the subsystem and cannot, in general, be answered completely by considering the conditional probability at either clock reading separately.  To address this question, it is useful to consider the density operator of the conditional states.

Although the clock states $|\beta_0\rangle$ and $|\beta_0'\rangle$ are orthogonal and therefore represent distinguishable clock readings, the global state may still contain quantum coherence between these different clock alternatives. Such coherence is encoded in the off-diagonal elements of the reduced clock density matrix and should arise from the quantum correlation between different clock states and the remaining subsystem. The magnitude of this correlation  should therefore provide information about the relation between the conditional states associated with different clock readings: a larger coherence indicates stronger quantum correlation between the corresponding relational descriptions, whereas lesser correlation indicates that such descriptions become increasingly weaker.

\subsection{Density Matrix from Conditional States}
The complete density operator for the global state~\eqref{global_state} 
is given by
\begin{equation}
\hat{\varrho} = |\Psi_N\rangle\langle\Psi_N|.
\label{rho1}
\end{equation}
Substituting from Eq.~\eqref{global_state} in~\eqref{rho1}, the expression for density operator turns out  to be
\begin{align}
\hat{\varrho} = \int d\alpha\,d\beta \int d\alpha'\,d\beta'\, \Psi_N(\alpha,\beta)
\Psi_N^*(\alpha',\beta') |\beta\rangle\langle\beta'| |\alpha\rangle_R\,{}_R\langle\alpha'|.
\end{align}

We can now construct the density operator associated with the conditional state~\eqref{condtn_state} as
\begin{equation}
\hat{\rho} = |\psi_{\beta_0}\rangle_R  {}_R\langle \psi_{\beta'_0}|.
\label{rho2}
\end{equation}
Substituting Eq.~\eqref{condtn_state} in~\eqref{rho2}, we obtain the conditional density operator,
\begin{align}
\hat{\rho} ={}& \int d\alpha\,d\alpha'\, \Psi_N(\alpha,\beta_0)
\Psi_N^*(\alpha',\beta'_0) |\alpha\rangle_R\,{}_R\langle\alpha'|.
\end{align}
Its matrix elements in the $\alpha$-basis are therefore
\begin{equation}
{}_R\langle\alpha| \hat{\rho} |\alpha'\rangle_R = \Psi_N(\alpha,\beta_0)
\Psi_N^*(\alpha',\beta'_0) = \rho(\alpha, \beta_0; \alpha', \beta'_0),
\label{rho3}
\end{equation}
upon using the orthonormality condition $\langle \alpha|\alpha' \rangle = \delta(\alpha-\alpha')$.

The reduced density matrix $\rho(\beta_0,\beta'_0)$ is obtained by tracing over the states $|\alpha \rangle$ of the remaining subsystem in~\eqref{rho3}, giving
\begin{align}
\rho(\beta_0,\beta'_0) = \int d\alpha\, \rho(\alpha, \beta_0; \alpha, \beta'_0) = \int d\alpha\,
\Psi_N(\alpha,\beta_0) \Psi_N^*(\alpha,\beta'_0).
\label{reduced_rho}
\end{align}

Using the the conditional wave function~\eqref{Norm_wavefunction} in equation~\eqref{reduced_rho}, the
reduced density matrix is obtained as
\begin{align}
\rho(\beta_0,\beta'_0) =  \frac{N^2\sigma^2}{4} e^{ik_0(\beta'_0-\beta_0)} &
\int_{-\infty}^{\infty}d\alpha\, e^{-\frac{\sigma^2}{2} \left[(\alpha-\beta_0)^2+(\alpha-\beta'_0)^2\right]} \left[ 1+ \erf \left( \frac{-k_0-i\sigma^2(\alpha-\beta_0)}
{\sigma\sqrt{2}} \right) \right] \nonumber\\ &\times \left[ 1+ \erf \left( \frac{-k_0+i\sigma^2(\alpha-\beta'_0)} {\sigma\sqrt{2}} \right)\right].
\label{rho_beta_integral}
\end{align}
To simplify the Gaussian part of the integrand, we write
\begin{equation}
(\alpha-\beta_0)^2 + (\alpha-\beta'_0)^2 = 2\left( \alpha-\frac{\beta_0+\beta'_0}{2}
\right)^2 + \left( \frac{\beta_0-\beta'_0}{2} \right)^2
\label{exp_simplify}
\end{equation}
and introduce the shifted integration variable $u= \alpha-\frac{\beta_0+\beta'_0}{2}$. Using the variable $u$,  we can write 
\begin{equation}
\alpha-\beta_0 = u+\frac{1}{2}(\beta'_0-\beta_0)\ , \qquad \alpha-\beta'_0
= u-\frac{1}{2}(\beta'_0-\beta_0).
\end{equation}
Therefore, the error-function arguments can then be written compactly as
\begin{align}
-\frac{k_0}{\sigma\sqrt{2}} \pm \frac{i\sigma}{\sqrt{2}}(\alpha-\beta_0) = A \pm i b u,
\label{erf_arg}
\end{align}
where $A=a-\frac{i\sigma}{2\sqrt{2}}(\beta'_0-\beta_0)$, with $a=-\frac{k_0} {\sigma\sqrt{2}}$ and  $b=\frac{\sigma}{\sqrt{2}}$.

Substituting Eq.~\eqref{exp_simplify} and~\eqref{erf_arg}  in Eq.~\eqref{rho_beta_integral}, we obtain the reduced density matrix in the form 
\begin{align}
\rho(\beta_0,\beta'_0) = & \frac{N^2\sigma^2}{4} e^{ik_0(\beta'_0-\beta_0)} e^{-\frac{\sigma^2}{4}(\beta'_0-\beta_0)^2}  \int_{-\infty}^{\infty} du\, e^{-\sigma^2u^2}
\left[ 1+\erf(A-ibu) \right] \left[ 1+\erf(A+ibu)\right].
\label{reduced_rhoint}
\end{align}
This integral has the same structure as equation~\eqref{integralI} whose solution is given by equation~\eqref{Ifinal} in appendix~\ref{norm_calc}. Following that result,  we obtain the reduced density matrix~\eqref{reduced_rhoint} as
\begin{equation}
\rho(\beta_0,\beta_0') = \frac{e^{ik_0(\beta_0'-\beta_0)} e^{-\frac{\sigma^2}{4}(\beta_0'-\beta_0)^2}}{
1+\erf \left(-\frac{k_0}{\sigma}\right)} \left[1+\erf \left(-\frac{k_0}{\sigma}-\frac{i\sigma}{2}(\beta_0'-\beta_0)\right)\right].
\label{reduced_density_matrix}
\end{equation}

The matrix element $\rho(\beta_0,\beta_0')$ measures the overlap between conditional states with two different clock readings, $\beta_0$ and $\beta_0'$. Therefore, the off-diagonal elements in the clock basis encode the coherence between different clock readings. In particular, a suppression of $\rho(\beta_0,\beta_0')$ with the increase in  $|\beta_0-\beta_0'|$ should indicate that the corresponding conditional states become increasingly uncorrelated.

\subsection{Clock Coherence in Page-Wootters Framework}
In order to study the coherence between different clock readings, it is useful to consider the modulous of the reduced density matrix~\eqref{reduced_density_matrix},
\begin{equation}
\mathscr{C}(\beta_0,\beta_0')=\left|\rho(\beta_0,\beta_0')\right|,
\label{coherence}
\end{equation}
which provides a measure of the degree of coherence between different clock readings, $\beta_0$ and $\beta_0'$.

Since the conditional states are normalized, we have $\rho(\beta_0, \beta_0) = 1 $. Consequently, the degree of coherence $\mathscr{C}(\beta_0,\beta_0')$ has the range
\begin{equation}
0\leq \mathscr{C}(\beta_0,\beta_0')\leq 1,
\end{equation}
$\mathscr{C}(\beta_0,\beta_0')=1$ corresponding to identical  conditional states. On the other hand, $\mathscr{C}(\beta_0,\beta_0')=0$ corresponds to  perfectly uncorrelated conditional states.

To investigate the behavior of the conditional states for different clock readings, we introduce the clock separation variable $\Delta\beta = \beta_0'-\beta_0$ in equation~\eqref{reduced_density_matrix} so that the degree of  coherence  given by~\eqref{coherence} assumes the form 
\begin{equation}
\mathscr{C}(\Delta\beta) = \frac{e^{-\frac{\sigma^2}{4}\Delta\beta^2}}{1+\erf\left(-\frac{k_0}{\sigma}\right)}\left|1+\erf\left(-\frac{k_0}{\sigma}-\frac{i \sigma \Delta\beta}{2}\right)\right|.
\label{eq:C_delta_beta}
\end{equation}
We use this expression to study the behaviour of coherence for small and large separations in clock readings.

\begin{figure}[t]
    \begin{center}
    {\includegraphics[width=0.85 \textwidth]{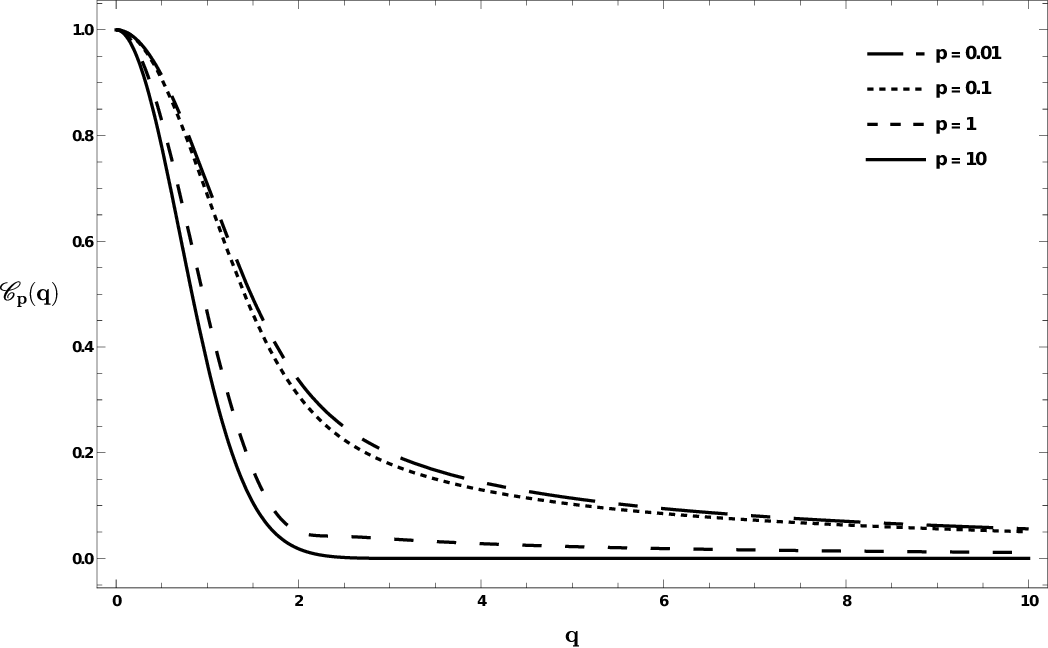}}\hfill%
     \end{center}
     \caption{\small \sf Behaviour of the coherence function $\mathscr{C}_p (q)$ expressed by equation~\eqref{C_xq} for the parameter values $p=0.01, 0.1, 1, 10$. For higher values of $p>10$, the curves almost coincide with the curve for $p =10$. On the other hand, for lower values of $p<0.01$, the curves almost coincide with the curve for $p =0.01$. This beahviour is discussed elaborately in the text.}
    \label{coherence-plot}
\end{figure}

It is useful to redefine  the parameters with $p=-\frac{k_0}{\sigma}$ and $q=\frac{\sigma\Delta\beta}{2}$, so that Equation~\eqref{eq:C_delta_beta} can be rewritten as
\begin{equation}
\mathscr{C}_p(q)= \frac{e^{-q^2}}{1+\erf(p)}\left|1+\erf(p-iq) \right|.
\label{C_xq}
\end{equation}
This coherence function is plotted with respect to $q =\frac{\sigma\Delta\beta}{2}$ for different values of $p$ in Figure~\ref{coherence-plot}.

In the next three subsections, we analytically assess the behaviour of the coherence function for different cases. 

\subsubsection{Behaviour of Clock Coherence}
We have obtained a general expression for the clock separation $\mathscr{C}_p(q)$ shown  by equation~\eqref{C_xq}. Figure~\ref{coherence-plot} displays the functional form of $\mathscr{C}_p(q)$ with respect to all clock separations $q = \frac{\sigma \Delta \beta }{2}$, for different values of the parameter $p=-\frac{k_0}{\sigma}$, namely, $p =0.01, 0.1, 1, 10$.

For higher values of $p>10$, the curves almost coincide with the curve for $p =10$. This is consistent with the fact that for $p \rightarrow \infty$, $\erf(p) \rightarrow 1$ and $\erf(p - i q) \rightarrow 1$,  so that the coherence function $\mathscr{C}_p(q)$ given by equation~\eqref{C_xq} approaches the form 
\begin{equation}
\mathscr{C}_p (q) \rightarrow e^{-q^2},
\end{equation}
a Gaussian curve, which is apparent in Figure~\ref{coherence-plot} for the case $p =10$.

On the other hand, for values of  $p <0.01$, the curves almost coincide with the curve for $p =0.01$.
This is due to the fact that for $p \rightarrow 0$, $\erf(p) \rightarrow 0$ and $\erf(p-i q) \rightarrow -i \erfi(q)$, so that the coherence function $\mathscr{C}_p(q)$ given by equation~\eqref{C_xq} approaches the form 
\begin{equation}
\mathscr{C}_p (q) \rightarrow e^{-q^2}\sqrt{1 +[\erfi(q)]^2}.
\end{equation}

The asymptotic behaviour $|1-\erf(iq)| = 1 + \frac{2 q^2}{\pi}+\ldots $  for small $q \rightarrow 0$ indicates the asymptotic expression
\begin{equation}
\mathscr{C}_p (q) \rightarrow e^{-q^2},
\end{equation}
a Gaussian in the region $q \sim 0$, which is apparent in Figure~\ref{coherence-plot}.

On the other hand, the asymptotic behaviour  $|1-\erf(i q)|= \frac{e^{q^2}}{\sqrt{\pi} q} (1 + \frac{1}{2 q^2}+ \ldots)$  for large $q \rightarrow \infty$ gives the asymptotic expression
\begin{equation}
\mathscr{C}_p (q) \rightarrow \frac{1}{\sqrt{\pi} q},
\end{equation}
a hyperbolic behaviour in the large $q$ tail-region, which is apparent in Figure~\ref{coherence-plot} for the case $p=0.01$.

\subsubsection{Clock Coherence for Proximal Clock Readings}
To study   small clock separations, we study the behaviour of the degree of coherence in the limit  $|\Delta\beta|\ll \frac{1}{\sigma}$.
For small $q$, we expand the error function in powers of $q$ using the Taylor series
\begin{equation}
\erf(p-iq) = \sum_{n=0}^{\infty} \frac{(-iq)^n}{n!} \frac{d^n}{dp^n}\erf(p).
\end{equation}
Since the first two derivatives of the error function are gicen by $\frac{d}{dp}\erf(p)=\frac{2}{\sqrt{\pi}}e^{-p^2}$ and $\frac{d^2}{dp^2}\erf(p) =-\frac{4p}{\sqrt{\pi}}e^{-p^2}$, we obtain
\begin{equation}
\erf(p-iq) = \erf(p) -\frac{2iq}{\sqrt{\pi}}e^{-p^2} +\frac{2 p q^2}{\sqrt{\pi}}e^{-p^2}
+\mathcal{O}(q^3).
\label{small_qexp}
\end{equation}
Let $C(p)=1+\erf(p)$ and $B(p)=\frac{2}{\sqrt{\pi}}e^{-p^2}$. Since  $p$ is real, $C(p)$ is positive and real-valued, and the small-$q$ expansion~\eqref{small_qexp} reads
\begin{equation}
1+\erf(p-iq) = C-iBq+p B q^2+ \mathcal{O}(q^3).
\end{equation}
Therefore, we have
\begin{equation}
\left|1+\erf(p-iq)\right| = C \sqrt{ 1+ \left( \frac{2 p B}{C} +\frac{B^2}{C^2}
\right)q^2 + \mathcal{O}(q^4) } =C + \left( p B +\frac{B^2}{2 C} \right)q^2 +\mathcal{O} (q^4),
\end{equation}
which can be re-expressed as 
\begin{equation}
\left|1+\erf(p-iq)\right| = 1+\erf(p) + \left[ \frac{2 p}{\sqrt{\pi}}e^{-p^2}
+ \frac{2e^{-2 p^2}} {\pi[1+\erf(p)]} \right]q^2 +\mathcal{O} (q^4).
\end{equation}
Since \(e^{-q^2}=1-q^2+O(q^4)\), Eq.~\eqref{C_xq} gives
\begin{equation}
\mathscr{C}_p(q) = 1- \left[1- \frac{2 p e^{-p^2}} {\sqrt{\pi}[1+\erf(p)]}
-\frac{2e^{-2 p^2}} {\pi[1+\erf(p)]^2} \right]q^2 + \mathcal{O}(q^4).
\end{equation}
Thus the coherence decreases quadratically with the clock separation $\Delta \beta = \frac{2 q}{\sigma}$,
\begin{equation}
\mathscr{C}_p(q) \simeq 1 - \mathscr{W} (p) q^2,
\label{C_xq_simplified}
\end{equation}
with no linear term in $q$.

The coefficient $\mathscr{W}(p)$ governing the leading-order decrement of  coherence is
\begin{equation}
\mathscr{W}(p) = 1- \frac{2 p e^{-p^2}} {\sqrt{\pi}\,[1+\erf(p)]} -\frac{2e^{-2 p^2}}
{\pi[1+\erf(p)]^2}.
\label{eq:Gamma_x}
\end{equation}
This coefficient depends only on the dimensionless parameter $p=-\frac{k_0}{\sigma}>0$. Thus, the leading-order behavior of the coherence is controlled by this ratio $\frac{k_0}{\sigma}$, rather than by $k_0$ and $\sigma$ independently.

The behavior of $\mathscr{W}(p)$ can be examined in two limiting regimes. For $\sigma \gg -k_0$, or equivalently $p \ll 1$,  we have $\erf(0) \approx 0$ so that $ \mathscr{W}(0) = 1-\frac{2}{\pi} \simeq 0.3634,$ yielding $ \mathscr{C}_p(q) \simeq 1 - 0.3634 \, q^2$. On the other hand, for $\sigma \ll -k_0$, or equivalently $p \gg 1$, we have $\erf(p)\to 1$ and $e^{-p^2}\to 0$, giving the estimate $\mathscr{W}(p) \to 1$ and consequently, $ \mathscr{C}_p(q) \simeq 1 - \, q^2$.
It follows that the degree of coherence decreases more rapidly in the case when $p \gg 1$ as compared to the case when $p \ll 1$ for a fixed  small clock separation $\Delta \beta$.

\subsubsection{Clock Coherence for Distant  Clock Readings}
For large clock separations, $q\gg 1$,  we use the asymptotic expansion for the complementary error function given in  Eq.~\eqref{erfc_asymptotic}, that is, $\erfc(z) = \frac{e^{-z^2}}{\sqrt{\pi} z}$ for $|z| \to \infty$.

Since $1+\erf(z)=2-\erfc(z),$ the coherence $\mathscr{C}_p(q)$ given  by Eq.~\eqref{C_xq} can be written as
\begin{equation}
\mathscr{C}_p(q) = \frac{e^{-q^2}}{1+\erf(p)} \left|2-\erfc(p-iq)\right|.
\label{C_xq_erfc}
\end{equation}
For $q\gg1$, the leading asymptotic contribution is
\begin{equation}
\erfc(p-iq)= \frac{e^{-(p-iq)^2}} {\sqrt{\pi}(p-iq)},
\end{equation}
which gives the modulous 
\begin{equation}
\left|\erfc(p-iq)\right| = \frac{e^{q^2-p^2}} {\sqrt{\pi}\sqrt{p^2+q^2}}.
\label{erfc_xq}
\end{equation}
Thus, for large separation of clock readings, $\Delta \beta = \frac{2 q}{\sigma}$,  we obtain 
\begin{equation}
\mathscr{C}_p(q) = \frac{1} {\sqrt{\pi}\,[1+\erf(p)]}\, \frac{e^{-p^2}}{\sqrt{p^2+q^2}}.
\label{eq:coherence_large_q}
\end{equation}

Equation~\eqref{eq:coherence_large_q} shows that the coherence decreases in a algebraic manner for large clock separations. Although the original large-$q$ expression contains a Gaussian factor $e^{-q^2}$, the exponential growth $e^{q^2}$ of the complementary error-function exactly cancels this Gaussian suppression in the asymptotic regime, giving rise to an algebraic decay.

In addition, for the case $p \ll 1$ (along with $q \gg 1$), we obtain a power law deacy with $\mathscr{C}_p(q) = \frac{1}{\sqrt{\pi} q}$. On the other hand, for the case $p \gg1$, the coherence is exponentially suppressed due to the factor $e^{-p^2}$.

\section{Discussion and Conclusion}
\label{discussion}
In this work, we investigated the fate of the big bang singularity in a plane-symmetric Bianchi I universe within the Wheeler-DeWitt framework of quantum cosmology. To address the absence of an external time parameter in the closed quantum system, we employed the Page-Wootters formalism, which provides a relational description of quantum evolution. This approach allowed us to formulate the dynamics in terms of conditional states and conditional probabilities, thereby enabling an examination of the big bang singularity from a relational perspective.

The model was formulated in a minisuperspace described by the Misner variables $\alpha$ and $\beta$, characterizing  the volume and anisotropy of the universe, respectively. The corresponding Hamiltonian constraint gives rise to the Wheeler-DeWitt equation,
$\hat{H}\Psi=(\hat{p}_\beta^2-\hat{p}_\alpha^2)\Psi=0.$
Since this equation has a Klein-Gordon-type structure, its natural inner product is not positive definite over the full space of solutions, making a straightforward probabilistic interpretation nontrivial. We therefore adopted the Dirac decomposition to recast the constraint into a first-order form, separating the physical solutions into positive- and negative-frequency sectors.

We decomposed the total Hamiltonian into the clock and the remaining degree of freedom, with $\hat{H}_C=\hat{p}_\beta$ and $\hat{H}_R=\sqrt{-\hat{p}_\alpha^2}$, respectively. This decomposition satisfies the essential commutation relations required in the Page-Wootters framework,
$[\hat{H}_C,\hat{H}_R]=0$ and $[\hat{\beta},\hat{H}_C]=i.$
The first relation ensures that the clock and the remaining subsystem are  noninteracting, while the second establishes $\beta$ as the variable canonically conjugate to the clock Hamiltonian.  Furthermore, the clock eigenstates form an orthogonal basis, $\langle\beta|\beta'\rangle=\delta(\beta-\beta'),$
so that distinct values of $\beta$ correspond to distinct clock readings. Together, these properties provide the kinematic and dynamical structure necessary to interpret $\beta$ as a relational clock for the evolution of the remaining subsystem.

The solution of the Wheeler-DeWitt equation is constructed as a superposition of momentum eigenstates, weighted by a Gaussian distribution function $F(k)$, centered around $k_0$ with width $\sigma$. We focused on the right-moving (expanding) branch of the universe, that corresponds to the negative-$k$
sector, $k<0$. The function $F(k)$ determines the spectral profile of the total state and brings about a correlation between the clock and the remaining subsystem. Consequently, the global state $|\Psi \rangle$ can be viewed as an entangled superposition of clock momentum states $|k\rangle_C$ and the states of the remaining subsystem, $|\psi_k\rangle_R$.

It is precisely this entanglement structure that underlies the emergence of relational dynamics in the Page-Wootters formalism. Conditioning the global state on a definite clock reading $\beta_0$ yields the corresponding conditional state of the remaining subsystem, thereby providing an effective notion of dynamics within the globally timeless framework. Accordingly, we obtain the conditional wavefunction associated with the clock reading $\beta_0$ by projecting the total state onto the clock eigenstate $|\beta_0\rangle$.

Since we have a Schr\"odinger-like equation following the Dirac decomposition,  we normalized the conditional state  using the standard Schr\"odinger inner product and defined the corresponding conditional probability density $P(\alpha|\beta_0)$ accordingly. Our analysis indicates that $P(\alpha| \beta_0) \to 0$ in the limit $\alpha \to -\infty$ for every finite value of $\beta_0$.
Since $\alpha =\frac{1}{3} \ln v$,  the limit $\alpha \to -\infty$ corresponds to vanishing volume, $v \to 0$, where the  big bang singularity occurs. Therefore, the vanishing of the conditional probability in the zero-volume limit indicates that the singular state is avoided in the quantum framework. Thus, within the Page-Wootters framework, the big bang singularity is effectively resolved at the level of conditional probabilities, yielding a nonsingular relational description of the quantum universe.

Beyond the construction of conditional states and the analysis of singularity avoidance, it is important to characterize how the conditional description of the remaining subsystem  changes between different clock readings.  This motivated considering the reduced density matrix $\rho(\beta_0, \beta_0')$ of the clock subsystem, obtained by tracing out the remaining degree of freedom $\alpha$. The  off-diagonal elements of $\rho(\beta_0, \beta_0')$ quantify the quantum coherence between different clock alternatives and therefore provide a measure of how strongly the corresponding relational descriptions remain quantum mechanically correlated. In particular, studying the dependence of these off-diagonal elements on the separation between two clock readings, $\Delta\beta=\beta_0'-\beta_0$, allowed us to investigate how the conditional states associated with different clock readings become progressively distinct within the global timeless quantum state.

The dependence of the coherence on the separation between clock readings was examined in two complementary regimes by introducing the quantities \(p=-k_0/\sigma\) and \(q=\sigma\Delta\beta/2\). For small clock separations, the coherence takes the form $ \mathscr{C}_p(q)\simeq 1-\mathscr{W}(p)q^2,$
where \(\mathscr{W}(p)\) increases from \(\mathscr{W}(0)=1-2/\pi\) and approaches unity as \(p\) becomes large. Thus, for nearby clock readings, the coherence remains close to its maximum value, with the leading loss of coherence being quadratic in the clock separation. At fixed \(q\), the increase of \(\mathscr{W}(p)\) with \(p\) indicates that coherence is lost more rapidly as \(p\) increases.

For large  clock separations, the coherence exhibits a qualitatively different asymptotic behavior. Using the large-$q$ limit, we obtained
$\mathscr{C}_p(q) = \frac{1} {\sqrt{\pi}\,[1+\erf(p)]}\, \frac{e^{-p^2}}{\sqrt{p^2+q^2}}$. 
This expression shows that the coherence decreases in an algebraic manner for large clock separations.   
In addition, for the case $p \ll 1$, this gives a power law deacy with $\mathscr{C}_p(q) = \frac{1}{\sqrt{\pi} q}$. On the other hand, for the case $p \gg1$, the coherence is exponentially suppressed due to the factor $e^{-p^2}$.

Taken together, these results reveal a nontrivial dependence of the temporal coherence on both the separation characteristic $q$ between clock readings and the spectral characteristics $p$ of the quantum state. For nearby clock readings, the coherence remains close to unity and decreases quadratically with the separation, whereas at sufficiently large separations its decay crosses over to an algebraic $1/q$ behavior.

Before we arrive at the conclusion, it is worthwhile to mention the notable highlights of the paper:

\begin{itemize}
    \item \textbf{Relational formulation of quantum cosmological
    dynamics:} We implemented the Page-Wootters framework to
    describe the evolution of the universe in terms of an internal
    clock, addressing the absence of an external time parameter
    in the Wheeler-DeWitt formulation.

    \item \textbf{Investigation of the big-bang singularity:}
    We derived the conditional probability density for the universe
    and showed that it vanishes in the zero-volume limit. This result
    indicates resolution of the classical big-bang singularity
   within this probabilistic framework.

    \item \textbf{Quantum coherence between distinct clock
    readings:} We constructed a density matrix using conditional
    states associated with different clock states and the
    remaining gravitational degree of freedom. By tracing out
    the latter, we obtained the reduced density matrix and
    characterized the coherence between distinct clock states.

    \item \textbf{Dependence of coherence on the spectral
    properties of the quantum state:} We analyzed the coherence
    as a function of clock separation and studied  how its
    behavior is governed by the spectral characteristics of the
    Gaussian wave packet. This provides a quantitative
    characterization of temporal coherence in the Page-Wootters
    description.
\end{itemize}
Thus, by examining the behavior at the big-bang singularity and coherence between clock states within a common relational framework, our study connects the probabilistic description of the early universe with  quantum correlations underlying the relational dynamics.

In summary, the study presented here provides a relational perspective on the resolution of the big bang singularity  and temporal coherence of the quantum state.  The analysis of the singular region shows how the quantum description modifies the classical notion of the big bang singularity, while the coherence demonstrates that the relation between different clock states contains nontrivial information about the underlying quantum state.

\section*{Appendices}
\appendix
\section{Wave Packet Construction}
\label{wavefunction-integral}
We constructed the wavepacket~\eqref{wavepacket} that satisfies the Wheeler-DeWitt equation  in the expanding branch of the universe. It involves the integral in the form
\begin{equation}
\Psi(\alpha,\beta) = \frac{1}{\sqrt{2\pi}} \int_{-\infty}^{0} dk\, \exp \left[
-\frac{(k-k_{0})^{2}}{2\sigma^{2}} + ik(\alpha-\beta) \right].
\label{wf1}
\end{equation}
To evaluate the integral, define $x = \alpha - \beta$ and  rewrite the exponent term as
\begin{align}
-\frac{(k-k_0)^2}{2\sigma^2}+ikx = -\frac{1}{2\sigma^2} \left[
k-\left(k_0+i\sigma^2x\right) \right]^2 +ik_0x -\frac{\sigma^2x^2}{2},
\end{align} 
and the wavefunction~\eqref{wf1} takes the form 
\begin{align}
\Psi(x) = \frac{1}{\sqrt{2\pi}} e^{ ik_0x -\frac{\sigma^2x^2}{2}}  \int_{-\infty}^{0} \exp \left[ -\frac{\left(k-k_0-i\sigma^2x\right)^2}
{2\sigma^2} \right]dk.
\label{wf2}
\end{align}
Introducing the variable $t = \frac{k-k_0-i\sigma^2x} {\sqrt2\,\sigma},$ the wavefunction becomes
\begin{equation}
\Psi(x) = \frac{\sigma}{\sqrt{\pi}} e^{ ik_0x -\frac{\sigma^2x^2}{2} }
\, \int_{-\infty}^{\frac{-k_0-i\sigma^2x}{\sqrt2\sigma}} e^{-t^2}\,dt.
\label{wf3}
\end{equation}
Using the definition of the error function,
\begin{equation}
\erf(z) = \frac{2}{\sqrt{\pi}} \int_{0}^{z} e^{-t^2}\,dt,
\end{equation}
the wavefunction in~\eqref{wf3} can be written in the form
\begin{equation}
\Psi(\alpha,\beta) = \frac{\sigma}{2} e^{i k_0 (\alpha-\beta)
- \frac{\sigma^2}{2}(\alpha-\beta)^2} \left[ 1 + \erf \left( \frac{-k_0 - i\sigma^2(\alpha-\beta)}{\sqrt{2}\sigma} \right) \right].
\end{equation}

\section{Normalization Constant}
\label{norm_calc}
To normalize the wavefunction,  we need to solve the integral~\eqref{normintegral}, given by
\begin{equation}
I(a) = \int_{-\infty}^{\infty} dx\, e^{-\sigma^2x^2} \left[1+\erf(a-ibx \right] \left[1+\erf(a+ibx)\right].
\label{integralI}
\end{equation}
Introducing the variable $y = bx$, we have  
\begin{equation}
I(a) = \frac{1}{b} \int_{-\infty}^{\infty} dy\, e^{-2y^2} \left[1+\erf(a-iy)\right] \left[1+\erf(a+iy)\right].
\label{integralI2}
\end{equation}

To evaluate $I$, we differentiate it with respect to the parameter $a$,
\begin{equation}
\frac{dI}{da} = \frac{2}{b\sqrt{\pi}} \int_{-\infty}^{\infty}dy\, e^{-2y^2}
\Big[ e^{-(a-iy)^2} \left[1+\erf(a+iy)\right] 
+ e^{-(a+iy)^2} \left[1+\erf(a-iy)\right] \Big].
\label{I_da}
\end{equation}
We rewrite equation~\eqref{I_da} as
\begin{equation}
\frac{dI}{da} = \frac{2e^{-a^2}}{b\sqrt{\pi}} \Big[J_+(a) + J_-(a)\Big],
\label{dIda}
\end{equation}
where we have introduced the auxillary integrals
\begin{equation}
J_{\pm}(a)= \int_{-\infty}^{+\infty} e^{-y^2 \pm 2iay} \left[1+\erf(a \pm iy)\right]dy.
\label{J_+1}
\end{equation}

\subsection*{Evaluation of the auxillary integrals}
To evaluate the two integrals given by equation~\eqref{J_+1}, we first consider the auxillary integral 
\begin{equation}
J_+(a) = \int_{-R}^{+R} e^{-y^2+2iay} \left[1+\erf(a+iy)\right]dy,
\label{J+}
\end{equation}
where we shall take the limit $R \to \infty$ at the end of calcualtion.

Using the identity 
\begin{equation}
\frac{d}{d z} \erf(z) = \frac{2}{\sqrt{\pi}} e^{- z^2},
\label{erf_diff}
\end{equation} 
and differentiating equation~\eqref{J+}, we get
\begin{align}
\frac{d J_+}{d a} =   2 i \int_{-R}^{+R}dy\, 
 \Big[ y e^{-y^2+2iay} \left[1+\erf(a+iy)\right] \Big] +   \frac{4 R}{\sqrt{\pi}} e^{-a^2}.
\label{dj+_1}
\end{align}
Furthermore, using the identity
\begin{equation}
 y e^{-y^2+2iay}  = i a e^{-y^2+2iay}  -\frac12 \frac{d}{dy} e^{-y^2+2iay},
\end{equation}
equation~\eqref{dj+_1} takes the form
\begin{align}
\frac{d J_+}{d a} =   2 i \int_{-R}^{+R}dy\, 
 \Bigg[ \Big( i a e^{-y^2+2iay}  -\frac12 \frac{d}{dy} e^{-y^2+2iay} \Big) \left[1+\erf(a+iy)\right] \Bigg] +   \frac{4 R}{\sqrt{\pi}} e^{-a^2}.
\label{dj+_2} 
\end{align}
Since the first term on the right hand side of equation~\eqref{dj+_2} gives back the auxiallry integral ~\eqref{J+}, we can rewrite equation~\eqref{dj+_2} as  
\begin{align}
\frac{d J_+}{d a} =  -2 a J_+ - i \int_{-R}^{+R}dy\, 
 \Bigg[ \Big(\frac{d}{dy} e^{-y^2+2iay} \Big) \left[1+\erf(a+iy)\right] \Bigg] +   \frac{4 R}{\sqrt{\pi}} e^{-a^2}.
\label{dj+_3} 
\end{align}
Carrying out the integration  by parts in equation~\eqref{dj+_3}, and making use of the identity~\eqref{erf_diff}, we get
\begin{align}
\frac{d J_+}{d a} =  -2 a J_+ -i  
& \Bigg[ \Big(   e^{-y^2+2iay} \Big) \left[1+\erf(a+iy)\right] \Bigg]_{-R}^{+R} \nonumber \\
& +i \int_{-R}^{+R}dy\,  \Bigg[ \Big(  e^{-y^2+2iay} \Big) \frac{2 i}{\sqrt{\pi}} e^{-(a+iy)^2} \Bigg]  +   \frac{4 R}{\sqrt{\pi}} e^{-a^2},
\end{align}
which reduces to
\begin{align}
\frac{d J_+}{d a} =  -2 a J_+ -i  
 \Bigg[ \Big(   e^{-y^2+2iay} \Big) \left[1+\erf(a+iy)\right] \Bigg]_{-R}^{+R} - \frac{4 R}{\sqrt{\pi}} e^{-a^2}   +   \frac{4 R}{\sqrt{\pi}} e^{-a^2}.
\end{align}
The last two terms in the above equation cancel each other,  and  the surface term vanishes in the limit $R \to \infty$. Consequently, we obtain
\begin{equation}
\frac{\partial J_+}{\partial a} + 2 a J_+ =0.
\label{J+diff}
\end{equation}
Equation~\eqref{J+diff} is a first-order differential equation, whose solution is immediatley obtained as
\begin{equation}
J_+(a)=C e^{-a^2},
\label{j_c}
\end{equation}
where $C$ is the integration constant.

To determine $C$, we evaluate $J_+(a)$  at $a=0$. From the definition of $J_+(a)$ given by Eq.~\eqref{J_+1}, we write
\begin{equation}
J_+(0) = \int_{-\infty}^{\infty} e^{-y^2} \left[1+\erf(iy)\right]dy.
\end{equation}
Since $\erf(iy) = i\,\erfi(y)$, and $\erfi(y)$ is an odd function, the contribution from the second term in the integral containing $\erf(iy)$ vanishes. Therefore,
\begin{equation}
J_+(0) = \int_{-\infty}^{\infty} e^{-y^2}\,dy = \sqrt{\pi}.
\end{equation}
Since $J_+(0)=C$ according to equation~\eqref{j_c}, we obtain $C=\sqrt{\pi}$. Hence, 
\begin{equation}
J_+(a)=\sqrt{\pi}\,e^{-a^2}.
\label{j+final}
\end{equation}

To evaluate $J_- (a)$, we proceed in a similar manner by defining another auxillary function 
\begin{equation}
J_-(a) = \int_{-R}^{+R} e^{-y^2-2iay} \left[1+\erf(a-iy)\right]dy.
\label{J-}
\end{equation}
Redoing similar steps as before, we find 
\begin{equation}
J_-(a)=\sqrt{\pi}\,e^{-a^2}.
\label{j-final}
\end{equation}

Substituting from equations~\eqref{j+final} and~\eqref{j-final} in~\eqref{dIda}, we obtain the differential equation
\begin{equation}
\frac{dI}{da} = \frac{4}{b}e^{-2a^2}.
\end{equation}
Integration with respect to $a$ yields
\begin{align}
I(a) = \frac{4}{b} \int e^{-2a^2}\,d a + C' = \frac{\sqrt{2 \pi}}{b}
\erf(\sqrt{2}a) + C',
\label{I(a)+C}
\end{align}
where $C'$ is the integration constant.

To determine $C'$,  we consider the limit $a \to \infty$ in the definition of $I(a)$ given by~\eqref{integralI2}. Using the property $\erf(z) \to 1$ for Re $z \to \infty$, we thus obtain
\begin{align}
I(\infty) = \frac{4}{b} \int_{-\infty}^{\infty} e^{-2 y^2}\,dy
= \frac{4}{b} \sqrt{\frac{\pi}{2}}.
\end{align}
Consequently, the intergation constant $C'$ in Eq.~\eqref{I(a)+C}  is obtained as $C'=\frac{\sqrt{2 \pi}}{b }$. 

Substituting this value of $C'$ in equation~\eqref{I(a)+C}, we obtain the required expression for the integral $I(a)$ as
\begin{equation}
I(a) = \frac{\sqrt{2 \pi}}{b} \left[ 1+\erf(\sqrt{2}a) \right].
\label{Ifinal}
\end{equation}
This integral is used to evaluate the normalization constant given by equation~\eqref{norm_constant} from equation~\eqref{normintegral} with the identifications $a=-\frac{k_0} {\sigma\sqrt{2}}$ and $b=\frac{\sigma}{\sqrt{2}}$.

\section{Conditional Probability at Early Times}
\label{probability_analysis}
We obtain the expression for conditional probability in equation~\eqref{P_Y_form}, which is
\begin{equation}
P(\alpha|\beta_0) = \frac{N^2 \sigma^2}{4} e^{-2Y^2} \left[1+\erf(z_-)\right] \left[1+\erf(z_+)\right] 
\label{P_Y_form2}
\end{equation}
where $z_\pm=a\pm iY$. 

We need its behaviour in the limit of early times, $\alpha \to -\infty$, which is equivalent to $|Y| \to \infty$.

Using $\erf(z)=1-\erfc(z)$, we write 
\begin{equation}
F =\left[1+\erf(z_-)\right] \left[1+\erf(z_+)\right]
\end{equation} 
as
\begin{align}
F = \left[2-\erfc(z_-)\right] \left[2-\erfc(z_+)\right] = 4-2\erfc(z_-)-2\erfc(z_+) +\erfc(z_-)\erfc(z_+).
\label{F_erfc_expanded}
\end{align}
Substituting the asymptotic expansion
\begin{equation}
\erfc(z) = \frac{e^{-z^2}}{\sqrt{\pi}\,z} \left[ 1-\frac{1}{2z^2} +\frac{3}{4z^4}
-\frac{15}{8z^6} +\cdots \right], \qquad |z|\rightarrow\infty.
\label{erfc_asymptotic2}
\end{equation}
for $z_\pm=a\pm iY$, we obtain
\begin{align}
F = \, & 4 -\frac{2e^{-z_-^2}}{\sqrt{\pi}\,z_-} \left( 1-\frac{1}{2z_-^2} +\frac{3}{4z_-^4} +\cdots \right)
-\frac{2e^{-z_+^2}}{\sqrt{\pi}\,z_+} \left( 1-\frac{1}{2z_+^2} +\frac{3}{4z_+^4}
+\cdots \right) \nonumber\\
&+ \frac{e^{-(z_-^2+z_+^2)}} {\pi z_-z_+} \left( 1-\frac{1}{2z_-^2} +\frac{3}{4z_-^4}
+\cdots \right) \left( 1-\frac{1}{2z_+^2} +\frac{3}{4z_+^4} +\cdots \right).
\label{F_asymptotic_expanded}
\end{align}

The quantities appearing in this expression simplify according to
\begin{align}
z_-^2 &=(a-iY)^2 =a^2-Y^2-2iaY, \nonumber \\
z_+^2 &=(a+iY)^2 =a^2-Y^2+2iaY, \nonumber \\
z_-z_+ &=(a-iY)(a+iY) =a^2+Y^2,  \nonumber \\
z_-^2+z_+^2 &=2(a^2-Y^2).
\label{zsum_square}
\end{align}
Therefore, $e^{-z_-^2} = e^{-a^2+Y^2}e^{2iaY},$ and $ e^{-z_+^2} = e^{-a^2+Y^2}e^{-2iaY}$. Multiplication by $\frac{N^2 \sigma^2}{4} e^{-2Y^2}$ in Eq.~\eqref{F_asymptotic_expanded}  gives the four contributions in the conditional probability~\eqref{P_Y_form2},
\begin{equation}
P(\alpha|\beta_0) = P_1 + P_2 +P_3+P_4.
\end{equation}
where
\begin{equation}
P_1 = N^2 \sigma^2 e^{-2Y^2},
\label{P1_asymptotic}
\end{equation}
\begin{equation}
P_2 = -\frac{N^2 \sigma^2}{2 \sqrt{\pi}} \frac{e^{-a^2-Y^2}e^{2iaY}} {a-iY}
\left( 1-\frac{1}{2(a-iY)^2} +\frac{3}{4(a-iY)^4} +\cdots \right),
\label{P2_asymptotic}
\end{equation}
\begin{equation}
P_3 = -\frac{N^2 \sigma^2}{2 \sqrt{\pi}} \frac{e^{-a^2-Y^2}e^{-2iaY}} {a+iY}
\left( 1-\frac{1}{2(a+iY)^2} +\frac{3}{4(a+iY)^4} +\cdots \right).
\label{P3_asymptotic}
\end{equation}
The last term requires a little more simplification. The exponential factor becomes $e^{-2Y^2}e^{-(z_-^2+z_+^2)} = e^{-2Y^2}e^{-2a^2+2Y^2} = e^{-2a^2}$. Hence,
\begin{align}
P_4 =\frac{N^2 \sigma^2 e^{-2a^2}} {4 \pi(a^2+Y^2)} \left( 1-\frac{1}{2(a-iY)^2}
+\frac{3}{4(a-iY)^4} +\cdots \right) \left( 1-\frac{1}{2(a+iY)^2} +\frac{3}{4(a+iY)^4} +\cdots \right).
\label{P4_before_product}
\end{align}
To leading orders, the product of the two conjugate series is
\begin{align}
&\left( 1-\frac{1}{2(a-iY)^2} +\frac{3}{4(a-iY)^4} +\cdots \right) \left( 1-\frac{1}{2(a+iY)^2} +\frac{3}{4(a+iY)^4} +\cdots \right) \nonumber\\
&\qquad = 1 -\frac{a^2-Y^2}{(a^2+Y^2)^2} +\frac{1}{4(a^2+Y^2)^2} +\frac{3(a^4-6a^2Y^2+Y^4)} {2(a^2+Y^2)^4} +\cdots .
\label{conjugate_series_product}
\end{align}
Thus,
\begin{align}
P_4 = \frac{N^2 \sigma^2 e^{-2a^2}} {4 \pi(a^2+Y^2)} \Bigg[ 1 -\frac{a^2-Y^2}{(a^2+Y^2)^2}
+\frac{1}{4(a^2+Y^2)^2} +\frac{3(a^4-6a^2Y^2+Y^4)} {2(a^2+Y^2)^4} +\cdots \Bigg].
\label{P4_asymptotic}
\end{align}
Therefore, the total expression for conditional probability in the limit $\alpha \to -\infty$ is given by summation of four terms thus evaluated, $P(\alpha|\beta_0) = P_1 + P_2 +P_3+P_4$.

\section*{Acknowledgements}
Vishal is supported through a Research Fellowship from the Ministry of Human Resource Development (MHRD), Government of India.


\end{document}